# Supplemental Material for Paper

## *Metallization of Nanofilms in Strong Adiabatic Electric Fields*

Maxim Durach[1], Anastasia Rusina[1], Matthias F. Kling[2], Mark I. Stockman[1, 2]

[1] Department of Physics and Astronomy, Georgia State University, Atlanta, GA 30303, USA

[2] Max-Planck-Institut für Quantenoptik, 85748 Garching, Germany

**CONTENTS**

## I. Crystal Model

Consider a cubic crystal formed by a periodic potential $V(\mathbf{r}) = U(x) + U(y) + U(z)$, where $U(x+a) = U(x)$ and $a$ is the lattice constant. The corresponding Hamiltonian eigenfunctions are the Bloch functions [1]

$$\psi_{M\mathbf{k}}(\mathbf{r}) = e^{i\mathbf{kr}} b_{M\mathbf{k}}(\mathbf{r}) = \phi_{mk_x}(x)\phi_{nk_y}(y)\phi_{qk_z}(z), \tag{1}$$

$$b_{M\mathbf{k}}(\mathbf{r}) = u_{mk_x}(x)u_{nk_y}(y)u_{qk_z}(z)\,;\ u_{mk}(x+a) = u_{mk}(x), \tag{2}$$

where $b_{M\mathbf{k}}(\mathbf{r})$ and $u_{mk}(x)$ are three- and one-dimensional Bloch amplitudes correspondingly, $M = (m, n, q)$ is the band index, and $\mathbf{k} = (k_x, k_y, k_z)$ is the quasi-wavevector of the state. The one-dimensional Bloch functions $\phi_{mk}(x) = e^{ikx}u_{mk}(x)$ are normalized as follows [2]

$$\int_{cell} u_{nk}^*(x)u_{mk}(x)dx = \frac{a}{2\pi}\delta_{mn} \text{ and } \int_{-\infty}^{\infty} \phi_{nq}^*(x)\phi_{mk}(x)dx = \delta_{mn}\delta(q-k). \tag{3}$$

The set of the Bloch functions is complete. The energy of the state (1) is given by

$$E_M(\mathbf{k}) = E_m(k_x) + E_n(k_y) + E_q(k_z), \tag{4}$$

where $E_n(k)$ is the one-dimensional Bloch electron dispersion in the *n*-th band. The dipole matrix element is equal to [2]

$$\int_{-\infty}^{\infty} \phi_{nq}^*(x)x\phi_{mk}(x)dx = -\delta_{mn}i\frac{\partial}{\partial q}\delta(q-k) + X_{mn}\delta(q-k), \text{ where} \tag{5}$$

$$X_{mn} = \frac{2\pi i}{a}\int_{cell} u_{nk}^*(x)\frac{\partial}{\partial k}u_{mk}(x)dx. \tag{6}$$

## II. Crystal in Electric Field. Transfer Matrix Approach

Consider an electric field $\mathcal{E}$ applied along the *x*-axis. The Schrödinger equation for motion of the particles with charge (-*e*) in the *x*-direction assumes the form:

$$H\varphi(x) = E\varphi(x), \text{ where } H = -\frac{\hbar^2}{2m_0}\frac{d^2}{dx^2} + U(x) + e\mathcal{E}x, \tag{7}$$

where $m_0$ is the bare electron mass, and potential $U(x) = -\alpha\sum_n \delta(x-na)$. Wave function $\varphi(x)$ in the interval $(n-1)a < x < na$ between the crystal planes is that of a free electron in electric field $\mathcal{E}$,

$$\varphi(x) = A_n \operatorname{Ai}\left(\frac{1}{\Lambda_0}\left(x - \frac{E}{e\mathcal{E}}\right)\right) + B_n \operatorname{Bi}\left(\frac{1}{\Lambda_0}\left(x - \frac{E}{e\mathcal{E}}\right)\right), \tag{8}$$

where $\Lambda_0 = \left(\frac{\hbar^2}{2m_0 e\mathcal{E}}\right)^{1/3}$ is the confinement length in electric field for bare electrons.

Transfer matrix $M$ for the transfer of the vector of coefficients through the period of the lattice in potential $U(x)$ is defined as

$$\begin{pmatrix} A_{n+1} \\ B_{n+1} \end{pmatrix} = M_n \begin{pmatrix} A_n \\ B_n \end{pmatrix}, \tag{9}$$

$$M_n = 1 + \frac{\pi\Lambda_0}{l_e}\begin{pmatrix} \operatorname{Ai}(Z_n)\operatorname{Bi}(Z_n) & \operatorname{Bi}^2(Z_n) \\ -\operatorname{Ai}^2(Z_n) & -\operatorname{Ai}(Z_n)\operatorname{Bi}(Z_n) \end{pmatrix}, \tag{10}$$

where 1 is the unity matrix, $l_e = \dfrac{\hbar^2}{2m_0\alpha}$ is an effective length of the periodic potential and

$$Z_n = \frac{1}{\Lambda_0}\left(na - \frac{E}{e\mathcal{E}}\right). \tag{11}$$

At the left boundary of the system $n = n_L = -N/2 + 1$, and at the right boundary $n = n_R = N/2$, where $N = L/2a$ is the number of the unit cells corresponding to the thickness $L$ of the nanofilm. The total transfer matrix $M$ between the boundary planes of the nanofilm is given by a product of the known partial transfer matrices $M_n$ as

$$M = \prod_{n=-N/2+1}^{n=N/2} M_n . \tag{12}$$

This transfer matrix can, in fact, be calculated completely analytically using *Mathematica* software (by Wolfram Research).

The coefficients of the wave function (8) at the left and right boundaries (denoted by the corresponding indices $A$ and $B$) are related (transferred) as

$$\begin{pmatrix} A_{n_R} \\ B_{n_R} \end{pmatrix} = M \begin{pmatrix} A_{n_L} \\ B_{n_L} \end{pmatrix}, \tag{13}$$

Assuming the electrons are confined inside the nanofilm by high enough potential walls, we impose zero boundary conditions $\varphi\left(\pm\frac{L}{2}\right) = 0$, which translate into

$$A_{n_L}\,\mathrm{Ai}\left(\frac{1}{\Lambda_0}\left(-\frac{L}{2} - \frac{E}{e\mathcal{E}}\right)\right) + B_{n_L}\,\mathrm{Bi}\left(\frac{1}{\Lambda_0}\left(-\frac{L}{2} - \frac{E}{e\mathcal{E}}\right)\right) = 0\,, \tag{14}$$

$$A_{n_R}\,\mathrm{Ai}\left(\frac{1}{\Lambda_0}\left(\frac{L}{2} - \frac{E}{e\mathcal{E}}\right)\right) + B_{n_R}\,\mathrm{Bi}\left(\frac{1}{\Lambda_0}\left(\frac{L}{2} - \frac{E}{e\mathcal{E}}\right)\right) = 0\,. \tag{15}$$

These equations (13)-(15) form a homogeneous linear system of four equations for four quantities $A_{n_L}$, $B_{n_L}$, $A_{n_R}$, and $B_{n_R}$. The condition of existence of its nontrivial solution is the vanishing of its determinant, which, in principle, allows one to find the energy spectrum. Under this condition, one of these quantities can be set arbitrarily. We set, e.g.,

$$A_{n_L} = 1, \quad B_{n_L} = -\frac{\mathrm{Ai}\left(\frac{1}{\Lambda_0}\left(-\frac{L}{2} - \frac{E}{e\mathcal{E}}\right)\right)}{\mathrm{Bi}\left(\frac{1}{\Lambda_0}\left(-\frac{L}{2} - \frac{E}{e\mathcal{E}}\right)\right)}, \tag{16}$$

thus satisfying Eq. (14). The normalization of the wave function at the end of the procedure will produce the standard wave function. The remaining Eqs. (13) and (15) depend on energy $E$ as a parameter. They can easily be resolved, resulting in a transcendental equation that we have to solve numerically. Note that except for this step, the rest of the problem is solved analytically, which dramatically increases numerical precision obtained at the end. Thus, in this case it is possible to reduce the Schrödinger equation, which is a partial-differential eigenproblem, to an explicitly analytical, though complicated, transcendental equation. As customary, we call this solution exact. (It is presently conventional to call analytical or exact a solution of a differential (eigen)problem that is reduced to explicit integrals, series, or a algebraic/transcendental equation. The entire solution, including their numerical evaluation, is often called semi-analytical.)

In realty, to calculate the energy spectrum of the system, we have employed the shooting method, which can be described as the following. We set the trial value of $E$ (this initial value should be negative enough). At the current step, we evaluate the transfer matrix of Eq. (12) and then coefficients $A_{n_R}$ and $B_{n_R}$ from Eq. (13). At the end of the current iteration, we compute the right-hand side of Eq. (15). Increasing the value of $E$, we find a point when this right-hand side changes sign, which correspond to the current eigenvalue of $E$. Then this procedure is repeated until required number of the eigenstates are evaluated. (This procedure reminds finding the right distance to target by gradually increasing the shooting angle, which gave it the name "shooting method".) With the spectrum of the system known, the wave functions are found analytically from Eq. (8).

## III. Crystal in Electric Field. Crystal Momentum Representation.

A solution of Eq. (7) in the crystal momentum representation can be written as an expansion over the Bloch functions:

$$\varphi(x) = \sum_m \int_{-\pi/a}^{\pi/a} dk\, a_m(k) \phi_{mk}(x) . \tag{17}$$

Substituting Eq. (17) into Eq. (7)

$$\sum_m \int_{-\pi/a}^{\pi/a} dk\, a_m(k)(E_m(k) - E + e\mathcal{E}\, x)\phi_{mk}(x) = 0 \tag{18}$$

and after multiplying by $\phi^*_{nq}(x)$ and integrating over $x$ we arrive at [3]

$$\begin{aligned} &(E_n(q) - E)a_n(q) + e\mathcal{E} \sum_m \int_{-\pi/a}^{\pi/a} dk\, a_m(k) \int_{-\infty}^{\infty} \phi^*_{nq}(x) x \phi_{mk}(x) dx = \\ &= \left( E_n(q) - E + ie\mathcal{E} \frac{\partial}{\partial q} \right) a_n(q) + e\mathcal{E} \sum_m a_m(q) X_{mn} = 0. \end{aligned} \tag{19}$$

In one-band approximation (i.e. $X_{mn} = 0$, if $m \neq n$) Eq. (19) has the form

$$\left( E_n(q) - \overline{E} + ie\mathcal{E} \frac{\partial}{\partial q} \right) a_n(q) = 0 \text{, where } \overline{E} = E - e\mathcal{E} X_{nn} \tag{20}$$

The solution to Eq. (20) is [3]

$$a_n(q) = A \exp\left\{ \frac{i}{e\mathcal{E}} \int_{q_0}^{q} (E_n(q') - \overline{E}) dq' \right\} \text{, where } q_0 \text{ is a constant,} \tag{21}$$

$$\varphi_n(x) = A \int_{-\pi/a}^{\pi/a} dk \exp\left\{ ik \left( x - \frac{\overline{E}}{e\mathcal{E}} \right) + \frac{i}{e\mathcal{E}} \int_0^k E_n(q) dq \right\} u_{nk}(x) . \tag{22}$$

## IV. Carriers in Region of Stopping Points

Consider eigenfunctions in the region of a stopping point $x \approx (\overline{E} - E_n)/e\mathcal{E}$ at the $n$-th band extrema at $k = \pm q_0$, where $E_n = E_n(\pm q_0)$,

$$\varphi_n(x) = A \int_{-\pi/a}^{\pi/a} dk \exp\left\{ ik \left( x - \frac{\overline{E} - E_n}{e\mathcal{E}} \right) + \frac{i}{e\mathcal{E}} \int_0^k (E_n(q) - E_n) dq \right\} u_{nk}(x) . \tag{23}$$

In the spirit of the stationary phase approximation the biggest contribution to the integral of Eq. (23) comes from the interval of $k$ where $(E_n(q) - E_n)/e\mathcal{E}a << 1$. Therefore, the Bloch electron dispersion under the integral can be expanded as $E_n(q) \approx E_n \pm \hbar^2 (q \pm q_0)^2 / 2m^*$, where $m^*$ is the effective mass at $k = \pm q_0$ (with "+" in front of the quadratic term in the dispersion corresponding to electrons and "-" to holes). As a result for $x \approx (\overline{E} - E_n)/e\mathcal{E}$

$$\varphi_n(x) \approx u_{nq_0}(x) \Psi_n(E, x) , \tag{24}$$

$$\Psi_n(E,x) = 2A\int_0^{\pi/a} dk \cos\left\{k\left(x - \frac{\overline{E}-E_n}{e\mathcal{E}}\right) \pm \frac{\hbar^2}{2m^* e\mathcal{E}}\frac{(k-q_0)^3}{3} + \frac{1}{e\mathcal{E}}\int_0^{q_0}(E_n(q)-E_n)dq\right\},$$

where $\Psi_n(E,x)$ is the envelope function. For the band extrema at the Γ-point $q_0 = 0$ the integral in Eq. (24) can be evaluated as

$$\begin{aligned}&\int_{-\pi/a}^{\pi/a} dk \exp\left\{ik\left(x - \frac{\overline{E}-E_n}{e\mathcal{E}}\right) \pm \frac{i\hbar^2}{2m^* e\mathcal{E}}\frac{k^3}{3}\right\} = \\ &= \pm\frac{1}{\Lambda}\int_{-\Lambda(\pi/a)}^{\Lambda(\pi/a)} d\xi \exp\left\{\pm\frac{i\xi}{\Lambda}\left(x - \frac{\overline{E}-E_n}{e\mathcal{E}}\right) + \frac{i\xi^3}{3}\right\} \approx N\,\mathrm{Ai}\left(\pm\frac{1}{\Lambda}\left(x - \frac{\overline{E}-E_n}{e\mathcal{E}}\right)\right),\end{aligned} \tag{25}$$

where $\Lambda = \left(\frac{\hbar^2}{2m^* e\mathcal{E}}\right)^{1/3}$ is the confinement length for a dressed carrier (electron or hole), and $C = const$. Thus, we have shown that the envelope function in the region of a stopping point at the Γ-point is given by the Airy function

$$\Psi_n(E,x) = N\,\mathrm{Ai}\left(\pm\frac{1}{\Lambda}\left(x - \frac{\overline{E}-E_n}{e\mathcal{E}}\right)\right), \tag{26}$$

where $N$ is a normalization constant.

## V. Quantum Bouncing of Bloch Electrons

Consider carriers that are confined between the walls of the film and Γ-point reflection position. Boundary condition at the wall $x = -L/2$ for conduction band quasi-electron envelope is $\Psi_c'(-L/2) = 0$, which results in condition $\frac{1}{\Lambda_e}\left(\frac{L}{2} + \frac{\overline{E}_i - E_{bc}}{e\mathcal{E}}\right) = \zeta_i$, where $\zeta_i$ is i-th root of $\mathrm{Ai}'(-x)$. The energy of the *i*-th state at the conduction band edge can be expressed as

$$\overline{E}_i = E_{cb} - e\mathcal{E}(L/2 - \zeta_i\Lambda_e). \tag{27}$$

The quasi-electron wave function can be written as $\varphi_{ni}(x) \approx N u_{n0}(x)\cdot \mathrm{Ai}\left(\frac{1}{\Lambda}\left(x + \frac{L}{2}\right) - \zeta_i\right)$. The normalization constant can be found following the approach of the Ref. [4]

$$\int_{-L/2}^{\infty} \mathrm{Ai}\left(\frac{1}{\Lambda}\left(x + \frac{L}{2}\right) - \zeta_i\right)^2 dx = \Lambda\zeta_i\,\mathrm{Ai}(-\zeta_i)^2\,. \tag{28}$$

Finally, as follows from Eqs. (26), (27) and (28) the normalized wave-function of the quasi-electron has the form

$$\varphi_{ci}(x) \approx \frac{1}{\Lambda_e^{1/2}\zeta_i^{1/2}\,\mathrm{Ai}(-\zeta_i)}\,\mathrm{Ai}\left(\frac{1}{\Lambda_e}\left(x-\frac{E_i-E_{bc}}{e\boldsymbol{\mathcal{E}}}\right)\right)u_{c0}(x)\,. \tag{29}$$

The intraband dipole matrix elements between the quantum bouncer quasi-electron states given by Eq. (29) can be also found following the recurrent approach of the Ref. [4]

$$\langle i|x|\,j\rangle = \delta_{ij}\left(-\frac{L}{2}+\frac{2}{3}\Lambda_e\zeta_i\right)+(1-\delta_{ij})\frac{(\zeta_i+\zeta_j)}{(\zeta_i\zeta_j)^{1/2}(\zeta_i-\zeta_j)^2}\Lambda_e\,. \tag{30}$$

Boundary condition for quasi-hole envelope at the wall $x = L/2$ is given by $\Psi_n(L/2)=0$, which leads to condition $\frac{1}{\Lambda}\left(\frac{L}{2}-\frac{\overline{E}_i-E_{vt}}{e\boldsymbol{\mathcal{E}}}\right)=\xi_i$, where $\xi_i$ is *i*-th root of $\mathrm{Ai}(-x)$. The energy of the states is

$$\overline{E}_i = E_{vt} + e\boldsymbol{\mathcal{E}}(L/2-\xi_i\Lambda_h)\,. \tag{31}$$

Normalization constant can be found from the integral [4]

$$\int_{-\infty}^{L/2}\mathrm{Ai}\left(-\frac{1}{\Lambda}\left(x-\frac{L}{2}\right)-\xi_i\right)^2 dx = \Lambda\,\mathrm{Ai}'(-\xi_i)^2\,. \tag{32}$$

As a result the normalized wave function of quasi-holes has the following form

$$\varphi_{vi}(x) \approx \frac{1}{\Lambda_h^{1/2}\,\mathrm{Ai}'(-\xi_i)}\,\mathrm{Ai}\left(-\frac{1}{\Lambda_h}\left(x-\frac{E_i-E_{vt}}{e\boldsymbol{\mathcal{E}}}\right)\right)\cdot u_{v0}(x)\,. \tag{33}$$

The dipole matrix element of the hole state in the quantum bouncer regime can be found as

$$\langle i|x|\,j\rangle = \delta_{ij}\left(\frac{L}{2}-\frac{2}{3}\Lambda_h\xi_i\right)+(1-\delta_{ij})\frac{2\Lambda_h}{(\xi_i-\xi_j)^2}\,. \tag{34}$$

From Eq. (27) the energy of the conduction band edge can be found as

$$E_b(\boldsymbol{\mathcal{E}}) = E_b - e\boldsymbol{\mathcal{E}}(L/2-\zeta_1\Lambda_e(\boldsymbol{\mathcal{E}}))\,. \tag{35}$$

The valence band edge, as follows from Eq. (31), is given by

$$E_t(\boldsymbol{\mathcal{E}}) = E_t + e\boldsymbol{\mathcal{E}}(L/2-\xi_1\Lambda_h(\boldsymbol{\mathcal{E}}))\,. \tag{36}$$

The band gap $E_g(\boldsymbol{\mathcal{E}})$ is thus equal to

$$E_g(\boldsymbol{\mathcal{E}}) = E_g - e\boldsymbol{\mathcal{E}}(L-\zeta_1\Lambda_e(\boldsymbol{\mathcal{E}})-\xi_1\Lambda_h(\boldsymbol{\mathcal{E}}))\,. \tag{37}$$

The metallization field $\mathcal{E}_m$ found from condition $E_g(\mathcal{E}_m)=0$ is

$$\mathcal{E}_m = \frac{E_g}{e[L-\zeta_1\Lambda_e(\mathcal{E}_m)-\xi_1\Lambda_h(\mathcal{E}_m)]}. \tag{38}$$

Since $L >> \Lambda_e(\mathcal{E}_m), \Lambda_h(\mathcal{E}_m)$ one can very well approximate expressions (35), (36), (37) and (38) by

$$E_{b,t}(\mathcal{E}) = E_{b,t} \mp e\mathcal{E}L/2,\ E_g(\mathcal{E}) = E_g - e\mathcal{E}L \text{ and } \mathcal{E}_m = E_g/eL \tag{39}$$

## VI. Wannier-Stark regime

If the state is confined between two reflections at stopping points it is disconnected from the walls. In this case the state is bulk-like [5]. Its normalized wave function is given by

$$\varphi_n(x) = \left(\frac{a}{2\pi}\right)^{1/2} \int_{-\pi/a}^{\pi/a} dk \exp\left\{ik\left(x-\frac{\overline{E}}{e\mathcal{E}}\right)+\frac{i}{e\mathcal{E}}\int_0^k E_n(q)dq\right\} u_{nk}(x). \tag{40}$$

The state is invariant with respect to transformation $x \to x+d,\ \overline{E} \to \overline{E}+e\mathcal{E}d$ . From the condition that the Bloch components at the Brillouin zone boundaries are equivalent, one obtains [3] the energies of the Wannier-Stark states

$$(-\overline{E}/e\mathcal{E})\frac{2\pi}{a}+\frac{1}{e\mathcal{E}}\int_{-\pi/a}^{\pi/a} E_n(q)dq = 2\pi l \tag{41}$$

$$\overline{E} = \frac{a}{2\pi}\int_{-\pi/a}^{\pi/a} E_n(q)dq - e\mathcal{E}al \tag{42}$$

We have found that the intraband dipole matrix elements between Wannier-Stark states are given by

$$\langle l|x|l'\rangle = -\frac{a}{2\pi e\mathcal{E}}\int_{-\pi/a}^{\pi/a} dk\, E_n(q)\exp\{ika(l-l')\},\ \text{for } l \neq l'. \tag{43}$$

## VII. Oscillator Strength and Dielectric Permittivity

Dielectric permittivity of the nanofilm is given by the following expression

$$\varepsilon(\omega) = 1+\frac{4\pi}{\hbar V}\sum_{if}\frac{(n_f^{3D}-n_i^{3D})|(d_x)_{if}|^2}{\omega_{if}-\omega-i0} = 1+\frac{4\pi}{\hbar La^2}\sum_{if}\frac{8(n_f-n_i)|(d_x)_{if}|^2}{\omega_{if}-\omega-i0}. \tag{44}$$

Here $n_j^{3D} = 8n_j$ denotes population of a three-dimensional state, while $n_j$ is population of a state in *x*-direction. The factor 8 originates from 2 occupied bands in y- and z- directions; another 2 is due to spin degeneracy. Also $(d_x)_{if} = e\langle f|x|i\rangle$ is a transition dipole moment corresponding to the transition frequency $\omega_{if}$. The imaginary part of the dielectric function is

$$\mathrm{Im}\varepsilon(\omega) = \frac{32\pi^2}{\hbar La^2}\sum_{if}(n_i - n_f)\left|(d_x)_{if}\right|^2 \delta(\omega_{if} - \omega)\,. \tag{45}$$

The combined oscillator strength in the low-frequency part of the spectrum is

$$\frac{f}{n} = \frac{ma}{\hbar L}\sum_j \omega_j \left|(x)_j\right|^2 , \tag{46}$$

where the sum is taken over the low-frequency transitions with transition frequencies $\omega_j$ and

$x_j = \langle f|x|i\rangle$.

The imaginary part of dielectric permittivity averaged over the low-frequency band $\Delta\omega$ is

$$\mathrm{Im}\bar{\varepsilon}(\omega) = \frac{32\pi^2 e^2}{\hbar\Delta\omega La^2}\sum_j \left|(x)_j\right|^2 \tag{47}$$

The static dielectric constant is given by

$$\mathrm{Re}\,\varepsilon(0) - 1 = \frac{64\pi e^2}{\hbar La^2}\sum_j \frac{\left|(x)_j\right|^2}{\omega_j} \tag{48}$$

## VIII. Additional Discussion: Breakdown, Possible Experimental Observations, and Some Technological Ramifications

In this Section, we briefly discuss the possibilities for experimental observations and technological ramifications in more detail than in the Letter text.

The metallization effect considered in the present Letter critically depends on the possibility to apply a rather high electric field $\mathcal{E}$ and observe the effect with or without damage (breakdown) to the crystal. The dielectric breakdown is a common name for a group of phenomena. The fundamental upper limit to the electric field that an insulator can withstand is imposed by the Zener tunneling, as described in the text of the Letter. There is also a number of phenomena called summarily as breakdown, which are due to acceleration of shallow-level electrons in the applied electric fields. Such shallow levels are produced by impurities (dopants), crystallographic defects, or may be surface states. Their acceleration by the applied electric field causes avalanche by impact ionization, resulting in the lattice damage.

The breakdown fields are maximum in high quality insulators crystals. In particular, the high quality insulator nanofilms used as the gate insulators in field-effect transistors have thickness $L \sim 1$ nm and work without a breakdown under potential difference $\Delta\varphi \lesssim 1$ V corresponds to the internal fields $\mathcal{E} \lesssim 0.1$ V/Å, which are just below the fields typical for the metallization threshold. It is actually somewhat likely that *the breakdown field* $\mathcal{E}_{BD} \sim 0.1$ V/Å *in the gate insulators is defined by the present metallization effect*.

As we will discuss below in this Section, in the optically-applied fields, the metallization causes the population of the conducting states and, consequently, high values of the dielectric permittivity of the film, making it effectively a metal. The dielectric screening by the conduction electrons lead to much lower fields inside than outside. This is responsible for the high fields tolerated by the metals at their surfaces, cf. Ref. [6]. The metallization causes a plethora of the well-known nanoplasmonic effects. In the strong fields, they can be observed with femtosecond resolution by various approaches of ultrafast nonlinear optics, including generation of harmonics [6], photoelectron emission [7-8], pump-probe spectroscopy, etc. In this case, even if the metalizing fields are strong enough to cause the breakdown and damage, the ultrafast (femto- and attosecond) measurements, in particular, the detection of harmonics or high-energy photoelectrons can be accomplished before the lattice disintegrates.

The theory in the present Letter is developed in terms of the electric field $\mathcal{E}$ that is the *internal* (inside the material) electric field. Correspondingly, the experimental manifestations and the possibilities to observe this effect crucially depend on the way this field is created. While there may exist many approaches to do so, we will discuss two basic ways: (i) by the electromagnetic wave incident from infinity (the scattering formulation) and (ii) by means of two conducting electrodes that form capacitor plates between which the nanolayer of the material is placed.

(i) This approach is typical for an experiment where an electromagnetic (optical, THz, microwave) wave is impingent on the nanolayer from a distant source in such a way that its polarization is normal to the nanolayer. In this case the field $\mathcal{E}$ of the wave inside the system is determined by the excitation field and the charges at the surfaces of the system induced by this field (the depolarization phenomenon). If the frequency of the field is far from any resonances of the system (which is true for a dielectric nanofilm or a metalized nanofilm) then the field at the surface is equal to the field $\mathcal{E}_\infty$ at the infinity (i.e., at the source) and the internal field is by a factor of $\varepsilon$ less, where $\varepsilon$ is the dielectric permittivity of the nanolayer, which should be computed including the metallization effect itself.

When the threshold field $\mathcal{E}_m$ of the metallization is reached, the dielectric permittivity increases more than tenfold to $\varepsilon \gtrsim 100$, which leads to $\mathcal{E} = \mathcal{E}_\infty/\varepsilon \ll \mathcal{E}_\infty$. This means a strong dielectric screening, which tends to keep the internal field at values just above $\mathcal{E}_m$. Such a metal-like dielectric screening protects the crystal from damage and allows for the external field to be quite large. For instance, in experiments [6] the source field intensity was $\mathcal{I}_\infty \sim 10^{11}\ \text{W/cm}^2$, while the system was resonant and the field at the surface had local intensity $\mathcal{I} \sim 10^{12}\text{-}10^{13}\ \text{W/cm}^2$. Such an intensity was not damaging for the metal of the nanostructure. Note that local intensity $\mathcal{I} \sim 10^{11}\ \text{W/cm}^2$ corresponds to field $\mathcal{E} = \sqrt{\frac{8\pi\mathcal{I}}{c}} \sim 0.1$ V/Å, which is a typical field for the metallization problem under consideration.

It may also be of interest to apply much larger fields, which will be damaging. However, for films with thickness $L\sim3$ nm, as shown in the Letter, the required excitation pulse is only a few femtoseconds. In such a case, the metallization will occur and can be probed by similar ultrashort pulses before the nanosystem is destroyed. Such a type of experiments is characteristic of optics of ultrafast and ultrastrong fields.

When expressed in terms of the *internal* field $\mathcal{E}$, the metallization problem is a purely quantum-mechanical problem, which is considered in the Letter. However, expressed in terms of the source field $\mathcal{E}_\infty$ it is a deeply nonlinear problem due to the dramatic change of $\varepsilon$ in the course of metallization. We will consider such a nonlinear problem elsewhere.

Small (nanoscopic) defects of the nanofilm surfaces in the case of the metallization will lead to the appearance of nanoplasmonic hot spots just as it occurs at the nano-inhomogeneities of the metal surfaces – cf. Ref. [9]. While they may cause ionization and breakdown, they are of interest as a phenomenon. In contrast, the variations of the thickness of the nanofilm under these excitation conditions will complicate effect but not lead to the breakdown. This is because the normal electric field in the nanofilm is $\mathcal{E}_\infty/\varepsilon$ irrespectively from the film thickness. Of course, sharp steps at the surfaces may create enhanced local fields as the defects do [9].

As we point out in the Letter, the large negative value of $\mathrm{Re}\,\varepsilon$ and significant dielectric losses described by $\mathrm{Im}\,\varepsilon$ are typical for plasmonic metals. The very existence the fields nanolocalized at the metal surfaces requires $\mathrm{Re}\,\varepsilon < 0$, which condition is satisfied for metalized nanofilms in a wide spectral area – see Fig. 3 (b)-(d). An important parameter in plasmonics is the so-called quality factor $Q = -\mathrm{Re}\,\varepsilon/_{\mathrm{Im}\,\varepsilon}$. For the metalized nanofilms, depending on the frequency and $\mathcal{E}$ this quality factor ranges $Q{\sim}1-10$, which is about the same as for gold in the visible spectral range.

(ii) The formulation of the metallization problem as an insulator crystal nanofilm between two conducting electrodes is that the defined *internal* field $\mathcal{E}$ is defined determined by ration of the external potential difference $\Delta U$ and thickness $L$. In such a case, the metallization, when developed may or may not lead to the breakdown depending on internal resistance of the source. If this resistance is low (in engineering terms, the source is a generator of potential difference), then the metallization will lead to very high currents and a damage of the crystal. Such a situation is typical for one of the most important potential applications of this theory: the supercapacitors. In such a case, the metallization is a defining factor for their energy capacity, which fundamentally limits the maximum field in the supercapacitor dielectric to the values $\mathcal{E} \sim 0.1$ V/Å, which by an order of magnitude lower than the limit imposed by the Zener breakdown.

Another important application of this formulation is the gate insulator in field-effect transistors. In this case, also the metallization will lead to the breakdown of this insulator and damage of the transistor. The metallization imposes here a fundamental limit on the gate field, which is a defining factor for the transistor performance. Note that actual fields in the transistor-gate insulators are on the same order as the metallization field $\mathcal{E}_m \sim 0.1$ V/Å. It is quite possible that the actual transistor-gate breakdown field $\mathcal{E}_{BD} \sim 0.1$ V/Å is defined by the metallization.

Considering, the variations of the nanofilm thickness, the metallization is defined by the potential drop across the nanofilm and not by the field $\mathcal{E}$ *per se*. This is evident the first of Eqs. (3) in the Letter, which states that the metallization field $\mathcal{E}_m$ is defined as $e\mathcal{E}_m L = E_g$, which implies that at the metallization point the total potential drop across the nanofilm $\Delta U = e\mathcal{E}_m L$ is fixed and equal to the bandgap $E_g$.

Therefore, in this case the variation in thickness will not lead to the breakdown in the thinnest part in contrast to other mechanisms of breakdown including the Zener tunneling and field-induced avalanche. However, the non-uniformity in the nanofilm thickness will lead to different limitations on the rate with field may change with the respect to the adiabaticity. Thinner films are more adiabatic and therefore more tolerant with respect to the metallization mechanism. One need also mention that the transistor (MOS) technology produces highly uniform oxide nanofilms used as the gate insulators, which will be very suitable for the purposes of studying the metallization effect.